# PHONON SPECTRUM AND VIBRATIONAL CHARACTERISTICS OF L INEAR NANOSTRUCTURES IN SOLID M ATRICES

E.V. Manzhelii[1], S.B. Feodosyev[1], I.A. Gospodarev[1], E.S. Syrkin[1], K.A. Minakova[2]

[1]B. Verkin Institute for Low Temperature Physics and Engineering, 47 Lenin Ave., 61103 Kharkov, Ukraine, emanzhelii@ilt.kharkov.ua

[2] National Technical University "Kharkiv Polytechnic Institute"

21, Frunze St., 61002, Kharkov, Ukraine

**Introduction:**

The relative simplicity of one-dimensional models often helps to solve the problem in an analytical form from the beginning to the end. For many years, one-dimensional models have retained attractive for a qualitative description of quasi-particle spectra and consequently physical characteristics of solids (see, e.g., [1]. Nondegeneracy of the energy eigenvalues and the absence of singularities on the spectral densities within the continuous spectrum band significantly simplify the problem of determining spectra perturbations induced by various defects [1, 2]. At once it should be noted that the formation of localized states in low-dimensional and three-dimensional systems are quite different (see, e. g., [3]). On the one hand, the use of one-dimensional models for the description of vibrations localized on defects in three-dimensional crystals reduces the value of this description. On the other hand, the zero-threshold forming of local discrete levels outside the quasi-continuous spectrum band and the strong dependence of energy levels on the defect parameters are of great general importance including applied purposes.

The in-stability of the crystalline phase of one-dimensional structures is the main disadvantage for the description of phonon spectra and vibrational characteristics of real systems. There is no long-range order in one-dimensional systems because even at $T = 0$ its mean-square displacements diverge [4]. However, if the phonon spectrum of a linear chain starts with a frequency different from zero, this divergence disappears and thus the linear chains can exist in reality and can be studied experimentally. They can exist as the linear chains of atoms in the bulk or on the surface of some solid matrices, i.e., quasi-chains of carbon atoms between graphene monolayers on silicon substrates [5] and chains of gold atoms deposited on the silicon surface [6-8]. The bundles of closed at the ends nanotubes enriched with gases containing quasi-one-dimensional nanoinclusions are of great interest. In these bundles, gas atoms are adsorbed mainly in the grooves between the nanotubes on the surface of bundles [9, 10]. Neutron diffraction data [9] and the behavior of the specific heat of these systems at low concentration of adsorbed gases [11-14] indicate the formation of quasi-one-dimensional structures by gas atoms.

In this paper, the phonon spectra of the linear chains of atoms on the surface or in the bulk of a crystal matrix and their mean-square displacements are calculated to determine the conditions and temperature ranges of stability of the crystalline phase of these chains. The behavior of low temperature heat capacity of the linear chains on the surface is analyzed.

**1. Linear Chain Applied to the Crystal Surface**

**1.1** *The model under consideration:*

Let us consider a linear chain applicated to the close-packed surface [111] of a semi-infinite fcc-crystal (Fig. 1).

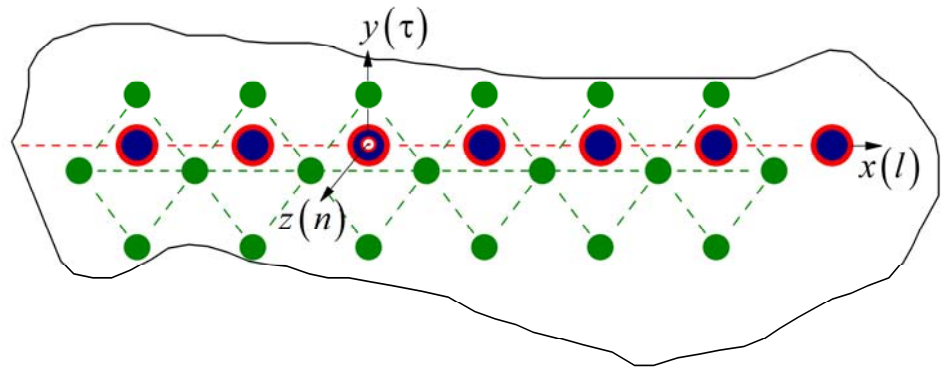

**Fig. 1**: The linear chain on the close-packed surface [111] of a semi-infinite fcc-crystal.

For simplicity, we assume that the interatomic interaction in the crystal-matrix is purely central, and its lattice constant $a$ coincides with the lattice constant of the applied chain. Further it will be shown that the crystal structure of the crystal matrix and the interactions between its atoms affect the spectrum of the chain very weakly.

The interaction of the chain with the crystal surface (substrate) we assume to be much weaker than the interaction between the atoms in the chain. In this case, it is natural to restrict ourselves to taking into account the nearest-neighbors interaction, i.e., each atom of the chain interacts with three atoms of the substrate. Therefore this interaction can also be considered to be purely central as nothing makes the chain to be at the distance $c$ from the surface, which corresponds to the minimum of the interaction potential between the atoms of the chain and the crystal-matrix ($c$ can differ from the interatomic distances in the chain and in the crystal). Therefore this interaction can also be considered to be purely central.

It is natural to describe the interaction between the atoms in the chain by the isotropic pair interaction potential $\varphi(\mathbf{\Delta}) = \varphi(\Delta)$ where $\mathbf{\Delta}$ is the vector connecting the interacting atoms. The force constant matrices have the form

$$\Phi_{ik}(\mathbf{\Delta}) = -\alpha(\Delta) \cdot \frac{\Delta_i \Delta_k}{\Delta^2} - \beta(\Delta) \cdot \delta_{ik},$$

Where the parameter $\beta(\Delta) \equiv \Delta^{-1} \cdot \partial\varphi/\partial\Delta$ describes the noncentral interaction between atoms at the distance $\Delta$ from each other, and the parameter $\alpha(\Delta) \equiv \partial^2\varphi/\partial\Delta^2 - \beta(\Delta)$ is the central interaction between these atoms.

Note that the weak interaction between atoms in the chain and in the substrate may be insufficient for the stability of the chain at vibrations in the direction along the plane and perpendicular to the chain (hereinafter it will be confirmed by calculations). Therefore, to ensure the stability of these vibrations, a flexural rigidity of the chain is required. The flexural rigidity is determined by the noncentral interaction of atoms in the chain, and for its description we have to use not only the nearest-neighbors interaction but also to take into account at least the nearest-neighbors interaction. The parameters describing this interaction $\beta_1 \equiv \beta = \beta(a)$ and $\beta_2 = \beta(2a)$ should fulfill the relation

$$\beta \cdot a^2 + \beta_2 \cdot (2a)^2 = 0. \qquad (2)$$

The relation (2) provides the transition of the lattice dynamics equations in the long-wave limit to the equations of the theory of elasticity (see, e.g., [1]) as well as the absence of the tension in the chain (see, e.g., [1]). By (2), the force constant matrices, which describe the interaction of the atoms in the chain with each other, can be written in the form

$$\Phi_{ik}(\pm a, 0, 0) = -(\alpha_1 \cdot \delta_{ix} + \beta) \cdot \delta_{ik}; \qquad (3)$$
$$\Phi_{ik}(\pm 2a, 0, 0) = -(\alpha_2 \cdot \delta_{ix} - \beta/4) \cdot \delta_{ik}.$$

In (3) $\alpha_1 \equiv \alpha(a)$ and $\alpha_2 \equiv \alpha(2a)$, the direction of the x-axis is chosen along the chain. The constant $\gamma_0$ denotes the force constant of the central interaction between atoms in the crystal-matrix. The central interaction between atoms in the chain with the nearest atoms in the crystal matrix is denoted by the force constant $\gamma$, and the difference between the distances $c$ and $a$ by the parameter $\delta \equiv \frac{c}{a}\sqrt{\frac{2}{3}}$. The force constant matrices describing these purely central interactions are of the form (1) at $\beta(\Delta) \equiv 0$. Then the self-interaction matrix of atoms in the chain [1] has the form

$$\Phi_{ik}(0,0,0) = -\sum_{\mathbf{\Delta}} \Phi_{ik}(\mathbf{\Delta}) =$$
$$= \left\{ \left[ 2(\alpha_1 + \alpha_2) + \frac{3}{2} \cdot \beta + \frac{3\gamma}{2(1+2\delta^2)} \right] \cdot \delta_{ix} + \left[ \frac{3}{2} \cdot \beta + \frac{3\gamma}{2(1+2\delta^2)} \right] \cdot \delta_{iy} + \left( \frac{3}{2} \cdot \beta + \frac{6\gamma \cdot \delta^2}{1+2\delta^2} \right) \cdot \delta_{iz} \right\} \cdot \delta_{ik} \qquad (4)$$

and it contains not only the parameters of the interaction between atoms in the chain but also the parameter of the chain-substrate interaction $\gamma$, which causes nonzero starting frequency of the chain spectrum.

**1.2 The spectral density of the atoms of the applicated chain with only nearest-neighbors interaction:** First we consider the deposited chain with only the nearest-neighbors interaction. It corresponds, for example, to the chains of solidified inert gases frequently used in recent years as the adsorbates (see, e.g., [11 15]). For these chains, as follows from (2), the interatomic interaction in the chain has to be purely central. In this case, the atomic vibrations normal to the chain are conditioned only by the interaction of the chain with the substrate (the surface of the crystal-matrix) and are localized. Hence the longitudinal vibrations have the dispersion law

$$\omega_l^2(\kappa) = \frac{2\gamma}{m} + \frac{4\alpha_1}{m} \cdot \sin^2 \frac{ak}{2}. \tag{5}$$

The quasi-wave vector $k$ can be introduced only for the crystallographic direction along the chain. Since this problem is purely model, for simplicity in (5) $\delta = 1$ is accepted.

From (5), for the phonon density of states of the chain we obtain

$$g_{ch}(\omega) = \frac{2}{\pi} \cdot \frac{\omega}{\sqrt{(\omega^2 - \omega_0^2)(\omega_m^2 + \omega_0^2 - \omega^2)}}, \tag{6}$$

here $\omega_0^2 = \gamma/2m$, $\omega_m^2 = 4\alpha_1/m$.

Figure 2 shows the spectral densities $\rho_l^{(ch)}(\omega)$, generated by the longitudinal displacement of the atoms the linear chain (curve 2), and obtained by using the method of Jacobian matrices [15, 17].

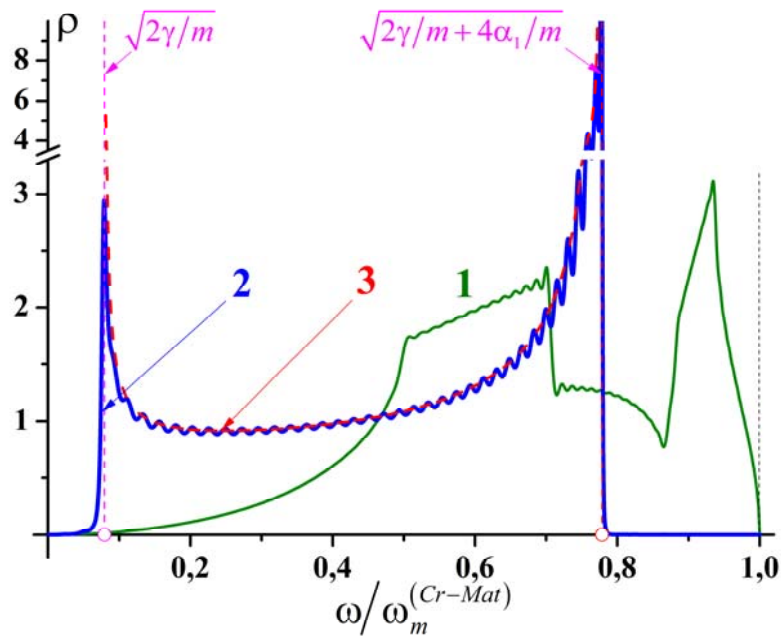

**Fig. 2**: The phonon density of the crystal-matrix (curve 1); the spectral density generated by the longitudinal displacement of an atom of the applicated chain (curve 2), and the phonon density of the linear chain in a periodic field (curve 3). The maximum frequency is $\omega_m^{(Cr-Mat)} = 4\sqrt{\gamma_0/m}$.

It is obvious that for a linear chain implanted into the surface layer or imbedded in the volume of the crystal-matrix, the behavior of the function $\rho_l^{(ch)}(\omega)$ is analogous to that of the deposited chain. The difference is that the starting frequency of the chain spectrum changes (for example, for a chain embedded in the volume, the initial frequency of the longitudinal oscillations is $2 \cdot \sqrt{\gamma/m}$ ).

Hence we can conclude that the longitudinal vibrations of the linear chain applied to the crystal surface or embedded in its volume do not propagate through the crystal-matrix and are completely localized on the chain. That is, both the spectrum and the structure of the crystal-matrix have virtually no effect on the spectral density of the chain. Thus the results of calculations of the spectral densities of linear chains deposited on the surface or embedded in the volume of the crystal with a very complex structure (e.g., silicon) are expected to coincide with those made for the simple model of the crystal-matrix.

### *1.3 The rms amplitudes of the vibrations and stability of the structure:*

Since the spectrum of the chain begins with a nonzero starting frequency, its longitudinal oscillations are stable, i.e., the rms displacements of atoms converge at any temperature. However, in the case of the chains applied to the surface, the amplitude of oscillations in the directions perpendicular to the chain can be large and even at low temperatures lead to the destruction of long-range order, namely, crystalline regularity of the atomic arrangement. Since, as noted above, there is every reason to assume that the interaction of the substrate with the chain is purely central (or, at least close to it (similar to it)), then the most "dangerous" are tangential oscillations, exactly the oscillations directed normal to the chain and along the plane of the crystal-matrix surface. The rms amplitudes (RMSA) of atomic displacements along various crystallographic directions (quantities $\sqrt{\langle u_i^2(r_{ch},T)\rangle}$ ) of the applicated chain are shown in Fig. 3.

It is seen that the RMSA of longitudinal atomic displacements in the chain are slightly higher than those of the crystal-matrix and in a wide range of temperatures is not a "threat" for the stability of the crystal. The same can be said about the RMSA of oscillations normal to the surface that exceed the RMSA of the longitudinal displacements at about 20%. When the next-nearest-neighbors interaction (both central and non-central) is taken into account, it slightly reduces the value of $\sqrt{\langle u_l^2(r_{ch},T)\rangle}$ and actually does not change the value $\sqrt{\langle u_n^2(r_{ch},T)\rangle}$. The RMSA of the tangential displacements are significantly higher than others. When only the nearest-neighbors interaction (curve $\tau 0$) is taken into account, then, even at low enough temperatures ($T \sim 0.2 \cdot \Theta_P$), the RMSA can lead to the destruction of the regularity of crystal structure. A consideration of the next nearest-neighbors interaction can greatly reduce the RMSA of the tangential displacements of the atoms in the chain (the curves $\tau 1$ and $\tau 2$ in Fig. 3 are significantly lower than the curve $\tau 0$). This reduction of the RMSA is determined by the flexural rigidity of the chain if the rigidity is positive. Taking into account (5), the dispersion laws of vibrations have the forms:

$$\omega_l^2(k) = \frac{\gamma}{2m} + \frac{4\alpha_1}{m} \cdot \sin^2\left(\frac{ka}{2}\right) + \frac{4\alpha_1}{m} \cdot \sin^2(ka) + \frac{4\beta}{m}\sin^4\left(\frac{ka}{2}\right); \qquad (7)$$

$$\omega_\tau^2(\kappa) = \frac{\gamma}{2m} + \frac{4\beta}{m}\sin^4\left(\frac{ka}{2}\right); \qquad (8)$$

$$\omega_n^2(\kappa) = \frac{2\gamma\delta^2}{m} + \frac{4\beta}{m}\sin^4\left(\frac{ka}{2}\right). \qquad (9)$$

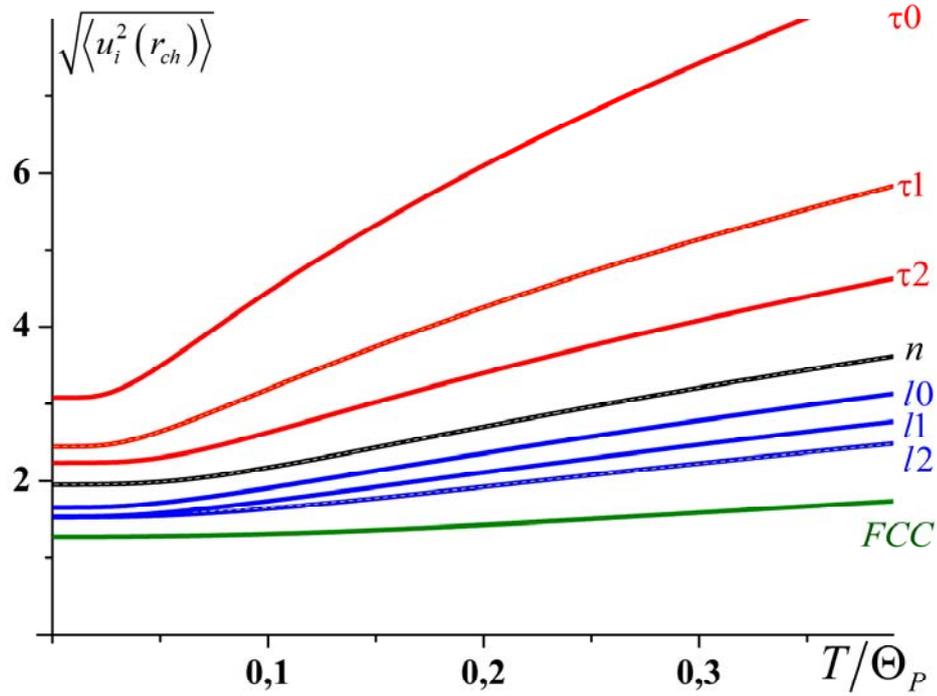

**Fig. 3:** Rms amplitudes of atomic displacements of the chain applicated to the surface [111] of fcc-crystal (here RMSA of fcc-crystal atoms are denoted by FCC). The curves $l0, l1, l2$ are RMSA of longitudinal displacements of the chain. The curves $\tau 0, \tau 1, \tau 2$ are RMSA of tangential displacements. Indices 0, 1, 2 indicate the increasing of flexural rigidity of the chain. Index 0 is zero flexural rigidity. $n$ denotes the RMSA of displacements normal to the crystal surface (their dependence on the flexural rigidity is weak). $\Theta_P \equiv \hbar\omega_0 / k$.

At low frequencies the contribution of non-central interaction in (7), (8), (9) is proportional to $\kappa^4$ and is negligible in (7). In (8) and (9), $\kappa^4$ determines the dispersion. At $\beta > 0$, (7), (8), (9) determine the flexural modes of the chain. Note that the positive values of $\beta$ correspond to the attraction between the atoms while the negative ones to the repulsion. Then the parameter $\beta > 0$ corresponds to the attraction of the nearest neighbors, and to the repulsion of the next-nearest-neighbors by (5). This is possible if the interaction between atoms at short and long distances has a different character and is described by different potentials. This interatomic interaction is inherent to carbon. In graphene and graphite, the nearest-neighbors interaction is covalent, and between more distant atoms it is Van der Waals'. For the second- and third-neighbors, the short-range covalent interaction is significantly small, and the Van der Waals interaction is strongly repulsive [18, 19], as the equilibrium distance of the corresponding potential is of the order of interlayer distance in graphite. This interatomic interaction causes a high flexural rigidity of graphene monolayers [20, 21]. Note that exactly carbon nanochains were studied in [5].

For one and two-dimensional structures of rare gas atoms the flexural rigidity is zero (or even insignificantly less than zero). Therefore, taking into account small values of $\Theta_P$ and melting temperature of the crystalline phase, the applicationof the linear chains of inert gases on the crystal surface is practically impossible. The linear chains implanted into the surface layer or the volume of the crystal are much more stable.

## 2. Linear chain implanted into the surface layer of the crystal:

*2.1. The model under consideration:* Let the linear chain be implanted into the surface layer of the fcc-crystal with purely central interaction between the atoms (see Fig. 4).

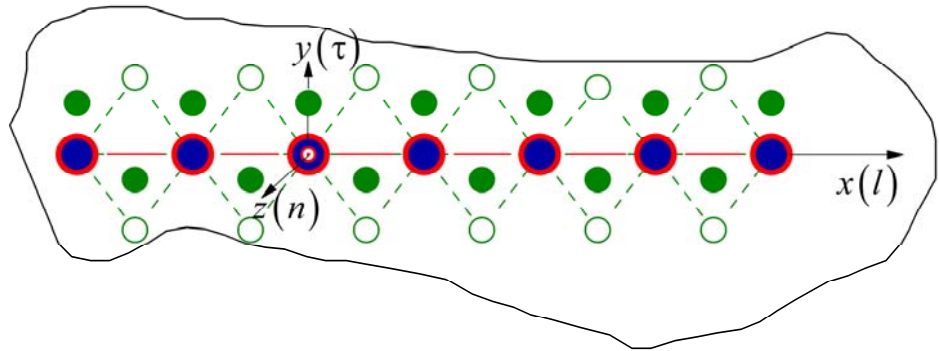

**Fig 4:** Linear chain implanted into the [111] surface of a semi-infinite fcc-crystal.

The interaction between the atoms in the chain and in the crystal-matrix can be considered to be purely central, as in the previous case, if the distances from the atoms in the chain to the atoms in the surface and subsurface layers coincide. This condition leads to the relation $c_s = c \pm \sqrt{c^2 - c_0^2}$, here $c$ is the distance from the chain to the subsurface layer, $c_s$ is the distance between the surface and the subsurface layers, $c_0 \equiv a\sqrt{2/3}$ is the distance between the planes [111] inside the crystal-matrix.

In the case under consideration the fulfillment of condition (2) is non-obligatory, because the implanted chain can be in a stressed state (the model considered in [22-24]). Just as in these papers, when describing the interaction between the atoms in the chain we restrict ourselves to the nearest-neighbors interaction. The corresponding force constant matrices have the form $\Phi_{ik}(\pm a, 0, 0) = -(\alpha \cdot \delta_{ix} + \beta) \cdot \delta_{ik}$, and the self-interaction matrix, which takes into account the ratio between the distances from the subsurface layer to the surface layer and to the chain can be

written as follows:

$$\Phi_{ik}(0,0,0) = -\sum_{\Delta} \Phi_{ik}(\Delta) =$$

$$= \left\{ \left[ 2(\alpha+\beta) + \frac{7\gamma}{2(1+2\delta^2)} \right] \cdot \delta_{ix} + \left[ 2\beta + \frac{19\gamma}{2(1+2\delta^2)} \right] \cdot \delta_{iy} + \left[ 2\beta + \frac{2\gamma(2\delta^2-1)}{1+2\delta^2} \right] \cdot \delta_{iz} \right\} \cdot \delta_{ik}, \quad (10)$$

where, $\delta \equiv c/c_0$, as in the previous case. That is, for different crystallographic directions the one-dimensional behavior of the phonon spectrum of the chain (analogous [22-24]) practically begins with the frequencies:

along the chain: $\omega_0^{(l)} = \frac{7\gamma}{2(1+2\delta^2) \cdot m}$;

along the normal to the surface: $\omega_0^{(n)} = \frac{2\gamma(2\delta^2-1)}{(1+2\delta^2) \cdot m}$;

normal to the chain and parallel to the surface layer plane: $\omega_0^{(\tau)} = \frac{19\gamma}{2(1+2\delta^2) \cdot m}$.

At $c_s > c$, the chain implanted into the surface layer may imitate the chain of atoms of inert gases adsorbed in the grooves between two nearest nanotubes on the surface of a bundle of carbon nanotubes. [11 14 22-25]. The parameters of the system are chosen to simulate the phonon spectrum of the chain of xenon atoms adsorbed in the grooves between the nanotubes, the frequencies being $\omega_0^{(n)} < \omega_0^{(l)} < \omega_0^{(\tau)}$. The spectral densities generated by displacements of the atoms of the chain along different crystallographic directions are shown in Fig. 5.

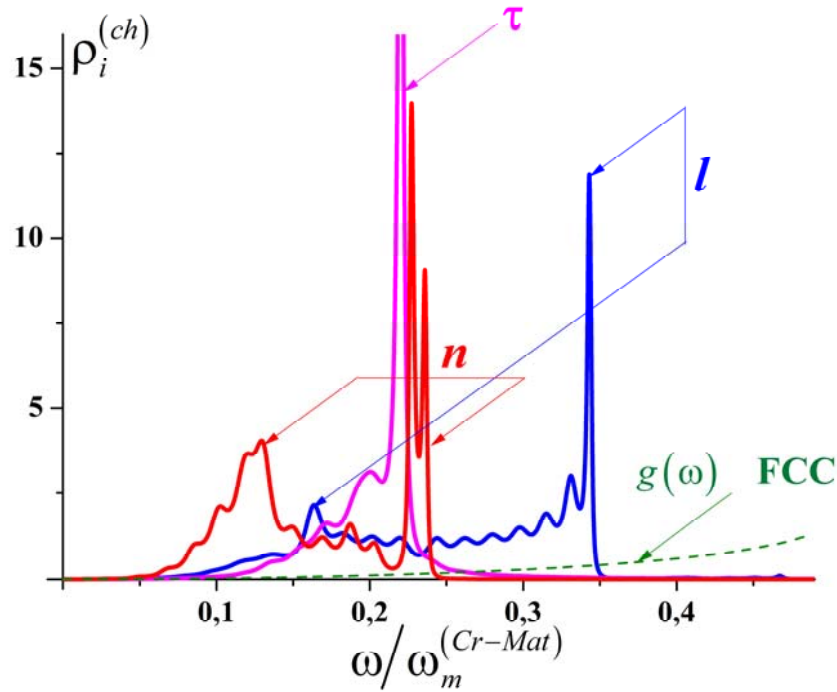

**Fig. 5.** The spectral densities generated by displacements of the atoms of the linear chain implanted into the surface of the semi-infinite fcc-crystal.

We can see that all the spectral densities are sufficiently different from zero, starting with a frequency somewhat lower than $\omega_0^{(n)}$. At lower frequencies the vibrations of atoms of the chain correspond to the acoustic vibrations of the total 3D system. Spectral densities of the atoms of the chain (except $\rho_n^{(ch)}(\omega)$) in this area are very small as a crystal-matrix is very rigid and $\omega_m^{(Cr-Mat)} \gg \omega_0^{(n)}, \sqrt{4\alpha_1/m}, \sqrt{4\beta/m}$. The spectral density $\rho_l^{(ch)}(\omega)$ generated by the longitudinal displacements of the atoms of the chain clearly shows quasi-one-dimensional behavior at $\omega > \omega_0^{(l)}$. This density differs from the similar density for the deposited chain considered in the previous section by higher values at $\omega_0^{(n)} < \omega < \omega_0^{(l)}$ and by an explicit three-dimensional behavior shown in this spectral interval. The spectral density $\rho_\tau^{(ch)}(\omega)$ generated by displacements of atoms in the direction perpendicular to the chain but parallel to the surface of the matrix crystal shows a three-dimensional behavior at $\omega_0^{(n)} < \omega < \omega_0^{(\tau)}$ and ends with a sharp resonance maximum at $\omega \approx \omega_0^{(\tau)}$, i. e., these vibrations practically do not propagate along the chain and do not interact with the longitudinal vibrations of the atoms. The vibrations of the atoms of the chain, polarized normally to the surface, show a three-dimensional behavior at $\omega < \omega_0^{(n)}$. At higher frequencies, the spectral density $\rho_n^{(ch)}(\omega)$ also shows a quasi-one-dimensional behavior, since thees vibrations hardly interact with longitudinal vibrations of the chain, and the vibrations in the direction $\tau$ are quasi-localized.

### *2.1. Heat capacity of implanted chains:*

It is well known that if the temperature tends to zero, then the phonon specific heat also tends to zero by a power law, the exponent being equal to the dimension of space. The linear behavior of the phonon specific heat at $T \to 0$ is a clear indication of one-dimensional behavior of the phonon spectrum. This, in turn, [4], results in a divergence of amplitude of atomic vibrations even at zero temperature, i.e., shows impossibility of these systems. The interaction with a three-dimensional matrix, shifting the beginning of the phonon spectrum of the chain to a value $\omega_0 > 0$, results in a change of the behavior of low-temperature heat capacity.

The phonon specific heat relates to the phonon spectrum (see, e.g., [1])

$$C_V(T) = qR \cdot \int \left(\frac{\hbar\omega}{2kT}\right)^2 \cdot \text{Sh}^{-2}\left(\frac{\hbar\omega}{2kT}\right) \cdot \nu(\omega) d\omega, \qquad (11)$$

where $q$ is the dimension of the space, and $\nu(\omega)$ is the phonon density of states of the system. Figure 6 shows the temperature dependence of the heat capacity calculated according to this formula with the phonon density of states (6) for different values $\omega_0/\omega_m$.

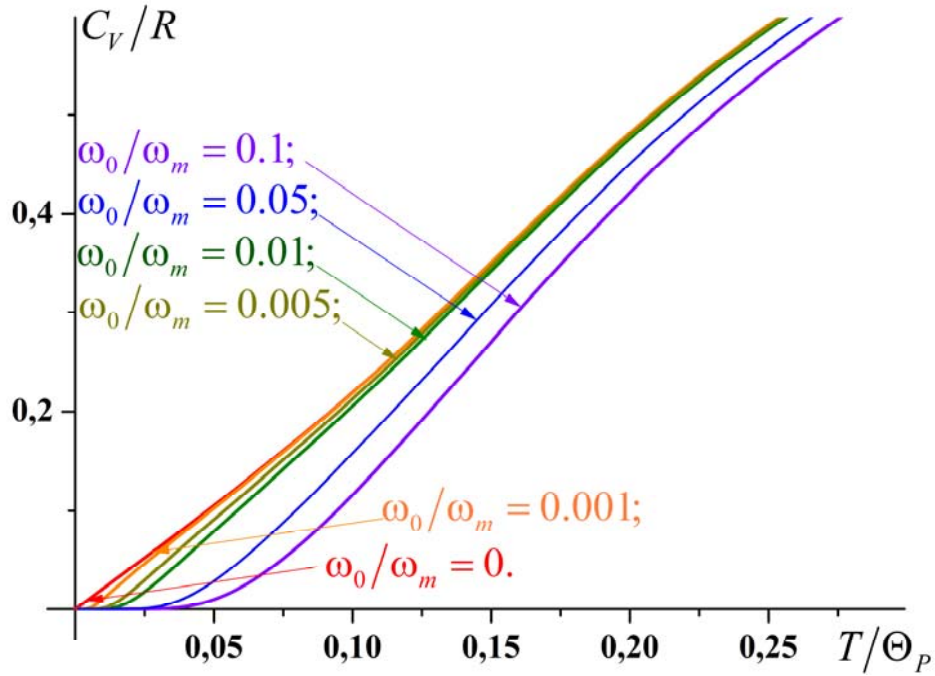

**Fig. 6**: Temperature dependence of the specific heat of linear chains in periodic external field.

It is seen clearly that, although there is a large linear region in all temperature dependences of heat capacity, at $T \to 0$ this dependence decreases exponentially. This linear region begins when the heat capacity begins to differ significantly from zero and ends at the inflection point on the dependence $C_V(T)$. The region of the exponential decrease of heat capacity is clearly visible even at $\omega_0/\omega_m = 0.001$. The linear behavior of the curve $C_V(T)$ at $T \to 0$ occurs only at $\omega_0 = 0$, which corresponds to an unstable system that does not really exist.

The model proposed in [22- 25] was based on the assumption that the interaction potential of atoms of the adsorbed chain with the walls of nanotubes does not depend on the coordinate along the direction of the tube, i.e., the tube is considered as a continuous medium. This assumption is invalid, because the distance between the chain and the wall of the tube is comparable with the distance between the carbon atoms in the nanotube itself. In addition, it should be noted that the chains of inert gas atoms adsorbed on the nanotube walls have large but still finite size (the number of atoms in the chain is $N \sim 10^4 \div 10^5$), and even in the free state the minimum natural frequency of this chain is $\omega_0 \sim 1/\sqrt{N} \leq 0.01$. That is, when $T \leq T^* \approx 0.2\Theta_P^{(ch)}$ ($\Theta_P^{(ch)} \equiv \hbar\omega_P^{(ch)}/k_B$,

$\omega_P^{(ch)}$ being the maximum frequency of oscillation of the chain), the heat capacity of the linear chains of inert gases adsorbed on the walls of nanotubes should decrease exponentially. The value $\Theta_P^{(ch)} \approx \Theta_D^{(G)}/2$, where $\Theta_D^{(G)}$ is the Debye temperature of a three-dimensional crystal of the corresponding solidified inert gas. The temperature dependence of the specific heat is linear at $T \geq T^*$.

Thus the effect of the periodic field of the crystal matrix (that provides the actual existence of a linear chain) on the low-temperature heat capacity can not be neglected.

Figure 7 shows the temperature dependence of the change of heat capacity of carbon nanotubes due to the adsorption of the chains of xenon atoms on them. The asterisks present the experimental curve obtained for nanotube bundles textured by compression [13]. The green curve corresponds to the heat capacity of the adsorbed chain calculated for the model proposed in [22-24], that is, a free infinite linear chain (an object that can not really exist). The violet curve shows microscopic calculations made for the model proposed in this section. It takes into account not only the heat capacity of the implanted chain (calculations made by formula (11) using the spectral densities shown in Fig. 5), but also the changed contributions of the atoms of the crystal matrix nearest to the chain to the heat capacity. On this curve, the exponential decrease of the change in heat capacity at $T \leq T^* \approx 1.5K$ is clearly shown, whereas on the curve calculated for the model [*] the specific heat at $T \to 0$ tends to zero linearly.

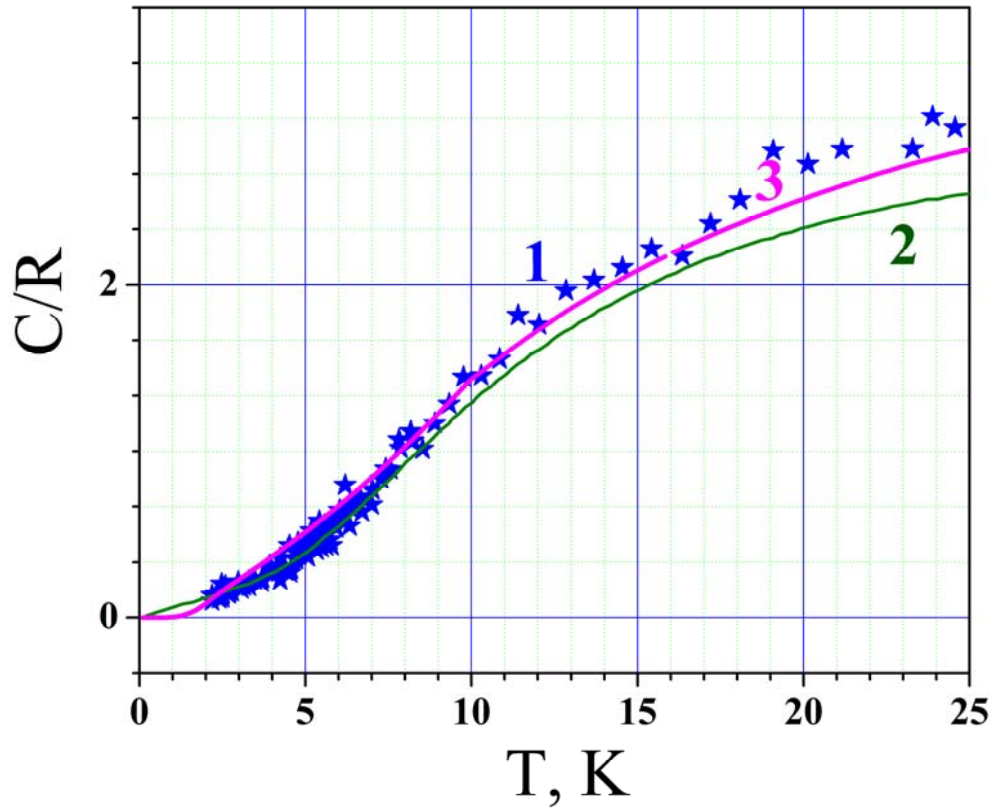

**Fig. 7:** Change in heat capacity of carbon nanotubes after the adsorption of xenon atoms.

Change in heat capacity, calculated in the frames of our microscopic model, is consistent with the experimental data, it is even slightly better than that calculated for the model [22-24]. However, non of calculations can explain the experiment [13] totally, because the contributions of all the degrees of freedom to heat capacity of nanotubes and the effect of adsorption of inert gas atoms to these contributions should be taken into account. Indeed, along with longitudinal vibrations of the atoms of carbon nanotubes that are responsible for an extremely high (~ 2500 K) value of their Debye temperature, their phonon spectra have low frequency flexural and torsional modes that are of quasi-one-dimensional type and may result in the dependence of $C_V(T)$ close to linear at low temperatures.

### Conclusion

Thus, after having calculated on a microscopic level the phonon spectra and fundamental vibrational thermodynamic characteristics of the linear chains applicated to a surface of the crystal matrix, or implanted into its volume or surface layer, we can conclude the following.

Starting with a frequency $\omega_0$, the vibrations of the linear chain applicated to the crystal surface or in the bulk, actually do not extend through the crystal matrix and are completely localized on the chain. The frequency $\omega_0$ is determined by the contribution of the weak interaction of an atom in the chain with the atoms of the crystal matrix in the self-interaction matrix of the atom in the chain. When $\omega > \omega_0$, neither the structure nor the phonon spectrum of the crystal matrix can hardly change the spectral characteristics of atoms in the chain.

At low frequencies ($\omega < \omega_0$), the vibrational spectrum of atoms of applied or implanted linear chains has a three-dimensional character. This determines the convergence of the mean-square displacements of atoms in the chain and the stability of these structures in a finite temperature range. Moreover, due to the three-dimensional behavior at low frequencies, the quasi-continuous character of spectrum is preserved in the presence of a finite concentration of defects in the chain.

At $\omega > \omega_0$, the vibrations of the chain atoms are either quasi-localized or their propagation has a one-dimensional character. The spectral densities of these atoms are well described by simple analytical expressions obtained for one-dimensional models. Quasi-continuity of the phonon spectrum of these systems is preserved in the presence of a finite concentration of defects. Excitations of the phonon spectrum generated by defects (including discrete impurity levels) can be calculated similarly to [26-29].

The fact that the one-dimensional behavior of the phonon spectrum of applied or implanted linear chains begins with a non-zero frequency significantly affects the low-temperature phonon specific heat, i.e., at $T \to 0$ there is necessarily a temperature range in which the dependence $C_V(T)$ decreases exponentially. This interval is evident even at very low frequencies $\omega_0$. Our calculations agree quite satisfactory with experimental data [13].

**References**

1. A. M. Kossevich, *The Crystal Lattice (Phonons, Solitons, Dislocations)* (WILEY-VCH Verlag Berlin GmBH, Berlin, 1999).
2. I.M. Lifshitz, *Nuovo Cim. Suppl.* **3**, 716 (1956).
3. Ландау Л.Д., Квантовая механика / Л.Д. Ландау, Е.М. Лифшиц .. М.: Наука, 1964. . . 768 с.
4. L.D. Landau, *JETP* **7**, 627 (1937).
5. . C. Jin, H. Lan, L. Peng, K. Suenaga, and S. Iijima, *Phys. Rev. Lett.* **102**, 205501 (2009).
6. Tsuyoshi Hasegawa, and Shigeyuki Hosoki, *Phys. Rev.* **B 54**, 10300 (1996).
7. . A. Kirakosian, R. Bennewitz, F.J. Himpsel, and L.W. Bruch, *Phys. Rev.* **B 67**, 205412 (2003).
8. G. Stan, M.J. Bojan, S. Curtarolo, S.M. Gatica, and M.W. Cole, *Phys. Rev.* **B 62**, 2173 (2000).
9. S.O. Diallo, B. F**a**k, M.A. Adams, O.E. Vilces, M.R .Johnson, H. Schober, and H.R. Clyde, *EPL* **88**, 56005 (2009).
10. J.C. Lasjaunias, K. Biljakovi**c**, J.L. Sauvajol, and P. Monceau, *Phys.Rev. Lett.* **91**, 025901 (2003).

11. M.I. Bagatskii, M.S. Barabashko, A.V. Dolbin, and V.V. Sumarokov, *Fizika Nizkikh Temperatur* **38**, 667–673 (2012) [*Low Temp. Phys.* **38**, 523 (2012)].
12. M.I. Bagatskii, M.S. Barabashko, V.V. Sumarokov, *Fizika Nizkikh Temperatur* **39** 568 (2013). [*Low Temp. Phys.* **39** 441 (2013)].
13. M.I. Bagatskii, V.G. Manzhelii, V.V. Sumarokov, and M.S. Barabashko, *Fizika Nizkikh Temperatur* **39**, 801 (2013).
14. M.I. Bagatskii, M.S. Barabashko, V.V. Sumarokov, *JETP Lett.* **99**, 532 (2014).
15. V. I. Peresada, Condensed Matter Physics in Russian, FTINT AN UkrSSR, Kharkov, 172, (1968).
16. V. I. Peresada, Doctoral Dissertation, Kharkov (1972).
17. V. I. Peresada, V. N. Afanas'ev, and V. S. Borovikov, *Fizika Nizkikh Temperatur* **1**, 461 (1975) [Sov. J. Low Temp. Phys. **1**, 227 (1975)].
18. I.A. Gospodarev, K.V. Kravchenko, E.S. Syrkin, and S.B. Feodosyev, *Fizika Nizkikh Temperatur* **35**, 751 (2009).
19. A Feher, I.A. Gospodarev, V.I. Grishaev, K.V. Kravchenko, E.S. Syrkin, and S.B. Feodosyev, *Fizika Nizkikh Temperatur* **35**, 862 (2009).
20. E.S. Syrkin, S.B. Feodosyev, K.V. Kravchenko, A.V. Yremenko, B.Ya. Kantor, and Yu.A. Kosevich, *Fizika Nizkikh Temperatur* **35**, 208 (2009)..
21. Alexander Feher, Eugen Syrkin, Sergey Feodosyev, Igor Gospodarev and Kirill Kravchenko, *Quasi-Particle Spectra on Substrate and Embedded Graphene Monolayers* , in "Physics and Applica-tion of Graphene – Theo-ry" (ed. S. Mikhailov) *"InTex" Open Access Publisher*, ISBN 978-953-307-996-7 Croatia, 2010
22. A. Šiber, *Phys. Rev.* **B 66**, 205406 (2002).
23. A. Šiber, *Phys. Rev.* **B 66**, 235414 (2002).
24. A. Šiber, *Phys. Rev.* **B 67**, 165426 (2003).
25. Milen K. Kostov, M. Mercedes Calbi,† and Milton W. Cole, *Phys. Rev.* **B 67**, 245403 (2003).
26. E.S. Syrkin, S.B. Feodosyev, *Fizika Nizkikh Temperatur* **20**, 586 (1994).
27. M.A. Mamalui, E.S. Syrkin, S.B. Feodosyev, *Fizika Nizkikh Temperatur* **24**, 773 (1998).
28. M.A. Mamalui, E.S. Syrkin, S.B. Feodosyev, *Fizika Nizkikh Temperatur* **25**, 72 (1999).
29. M.A. Mamalui, E.S. Syrkin, S.B. Feodosyev, *Fizika Nizkikh Temperatur* **25**, 976 (1999).